\documentclass[API,prl,preprint,superscriptaddress]{revtex4}
\usepackage{graphicx}

\begin{document}



\title{Universal response of the type-II Weyl semimetals phase diagram}

\author{P. R\"{u}{\ss}mann}
	\affiliation{Peter Gr\"{u}nberg Institut and Institute for Advanced Simulation, 
	Forschungszentrum J\"ulich and JARA, 52425 J\"ulich, Germany}
\author{A. P. Weber} 
	\affiliation{Institute of Physics, {\'E}cole Polytechnique F{\'e}d{\'e}rale de Lausanne, 1015 Lausanne, Switzerland}
	\affiliation{Swiss Light Source, Paul Scherrer Institute, 5232 Villigen-PSI, Switzerland}
\author{F. Glott} 
	\affiliation{Physikalisches Institut, Experimentelle Physik II, 
	Universit\"{a}t W\"{u}rzburg, Am Hubland, D-97074 W\"{u}rzburg, Germany}	
\author{N. Xu}
       \affiliation{Institute of Physics, {\'E}cole Polytechnique F{\'e}d{\'e}rale de Lausanne, 1015 Lausanne, Switzerland}
      \affiliation{Swiss Light Source, Paul Scherrer Institute, 5232 Villigen-PSI, Switzerland}
\author{M. Fanciulli}
       \affiliation{Institute of Physics, {\'E}cole Polytechnique F{\'e}d{\'e}rale de Lausanne, 1015 Lausanne, Switzerland}
      \affiliation{Swiss Light Source, Paul Scherrer Institute, 5232 Villigen-PSI, Switzerland}
\author{S. Muff}
       \affiliation{Institute of Physics, {\'E}cole Polytechnique F{\'e}d{\'e}rale de Lausanne, 1015 Lausanne, Switzerland}
       \affiliation{Swiss Light Source, Paul Scherrer Institute, 5232 Villigen-PSI, Switzerland}
\author{A. Magrez}
       \affiliation{Institute of Physics, {\'E}cole Polytechnique F{\'e}d{\'e}rale de Lausanne, 1015 Lausanne, Switzerland}
\author{P. Bugnon}
       \affiliation{Institute of Physics, {\'E}cole Polytechnique F{\'e}d{\'e}rale de Lausanne, 1015 Lausanne, Switzerland}
\author{H. Berger}
       \affiliation{Institute of Physics, {\'E}cole Polytechnique F{\'e}d{\'e}rale de Lausanne, 1015 Lausanne, Switzerland}     
\author{M. Bode} 
	\affiliation{Physikalisches Institut, Experimentelle Physik II, 
	Universit\"{a}t W\"{u}rzburg, Am Hubland, D-97074 W\"{u}rzburg, Germany}
\author{J. H. Dil}
       \affiliation{Institute of Physics, {\'E}cole Polytechnique F{\'e}d{\'e}rale de Lausanne, 1015 Lausanne, Switzerland}
       \affiliation{Swiss Light Source, Paul Scherrer Institute, 5232 Villigen-PSI, Switzerland}
\author{S. Bl\"{u}gel} 
\affiliation{Peter Gr\"{u}nberg Institut and Institute for Advanced Simulation, 
	Forschungszentrum J\"ulich and JARA, 52425 J\"ulich, Germany}
\author{P. Mavropoulos}         
      	\affiliation{Peter Gr\"{u}nberg Institut and Institute for Advanced Simulation, 
	Forschungszentrum J\"ulich and JARA, 52425 J\"ulich, Germany}
\author{P. Sessi} 
	\email[corresponding author: ]{sessi@physik.uni-wuerzburg.de}
	\affiliation{Physikalisches Institut, Experimentelle Physik II, 
	Universit\"{a}t W\"{u}rzburg, Am Hubland, D-97074 W\"{u}rzburg, Germany}

\date{\today}

\vspace{1cm}
\begin{abstract}

The discovery of Weyl semimetals represents a significant advance in topological band theory. 
They paradigmatically enlarged the classification of topological materials to gapless systems 
while simultaneously providing experimental evidence for the long-sought Weyl fermions. 
Beyond fundamental relevance, their high mobility, strong magnetoresistance, 
and the possible existence of even more exotic effects, such as the chiral anomaly, 
make Weyl semimetals a promising platform to develop radically new technology. 
Fully exploiting their potential requires going beyond the mere identification of materials 
and calls for a detailed characterization of their functional response, 
which is severely complicated by the coexistence of surface- and bulk-derived topologically protected quasiparticles, 
i.e., Fermi arcs and Weyl points, respectively. 
Here, we focus on the type-II Weyl semimetal class 
where we find a stoichiometry-dependent phase transition from a trivial to a non-trivial regime.  
By exploring the two extreme cases of the phase diagram, we demonstrate the existence 
of a universal response of both surface and bulk states to perturbations. 
We show that quasi-particle interference patterns originate from scattering events among surface arcs.   
Analysis reveals that topologically non-trivial contributions are strongly suppressed by spin texture. 
We also show that scattering at localized impurities generate defect-induced quasiparticles sitting close to the Weyl point energy. These give rise to strong peaks in the local density of states, which lift the Weyl node significantly altering the pristine low-energy Weyl spectrum. Visualizing the microscopic response to scattering has important consequences for understanding the unusual transport properties of this class of materials. Overall, our observations provide a unifying picture of the Weyl phase diagram.
\noindent
\end{abstract}

\pacs{}

\maketitle



The discovery of topological insulators (TIs) \cite{RevModPhys.82.3045,RevModPhys.83.1057} 
led to intense research efforts in searching for materials whose properties are determined by band structure topology. 
In this context, the recent discovery of Weyl semimetals represents a milestone \cite{Xu613, PhysRevX.5.031013}. 
Contrary to TIs, where topological properties only are only manifested at boundaries as gapless Dirac states, 
Weyl semimetals host bulk as well as surface topologically protected quasiparticles. 
In particular, Weyl points appear at crossing points that are protected by the topology of the bulk band structure. 
Because of the well-known surface-to-bulk correspondence, they are necessarily associated to the appearance 
of new topologically protected boundary modes at the surface, the so-called topological Fermi arcs, 
which occupy unclosed Fermi contours connecting Weyl points of opposite chirality \cite{PhysRevX.5.011029,HXB2015}.

Beyond their fundamental relevance, Weyl semimetals are characterized by intriguing transport properties. 
Very high mobility, extremely strong magnetoresistence \cite{SNS2015}, 
and even more exotic phenomena such as the chiral anomaly \cite{PhysRevX.5.031023} 
or the possibility for Fermi arcs to tunnel through the bulk via the Weyl points have been discussed \cite{MNH2016}.  
Because of the robustness inherited from topological protection, these effects are raising great expectations 
for direct applications of these materials in spintronics and magneto-electrics.

Weyl semimetals have been first experimentally discovered in the TaAs monopnictide family, 
known as type-I \cite{Xu613,PhysRevX.5.031013,YLS2015,XAB2015,LYS2016,LXW2015,PhysRevLett.115.217601}. 
More recently, it has been suggested that once Lorentz invariance is broken, 
a new flavor of Weyl material can be realized, the so-called type-II, 
where strongly tilted Weyl cones emerge at the boundaries between bulk electron and hole pockets \cite{SGW2015}. 
Theoretical predictions identified the Mo$_x$W$_{1-x}$Te$_2$ family as promising compounds 
\cite{SGW2015,PhysRevB.92.161107,PhysRevLett.117.056805,CXC2016}. 
One of the most interesting aspects of type-II Weyl materials is the possibility 
to continuously tune their topological properties by acting onto their stoichiometry \cite{CXC2016}. 
The resulting phase diagram offers an ideal platform to explore the functional response 
of the electronic properties and the existence of unifying trends within the Weyl phase. 

\begin{figure*}[t]   
	\includegraphics[width=0.98\textwidth]{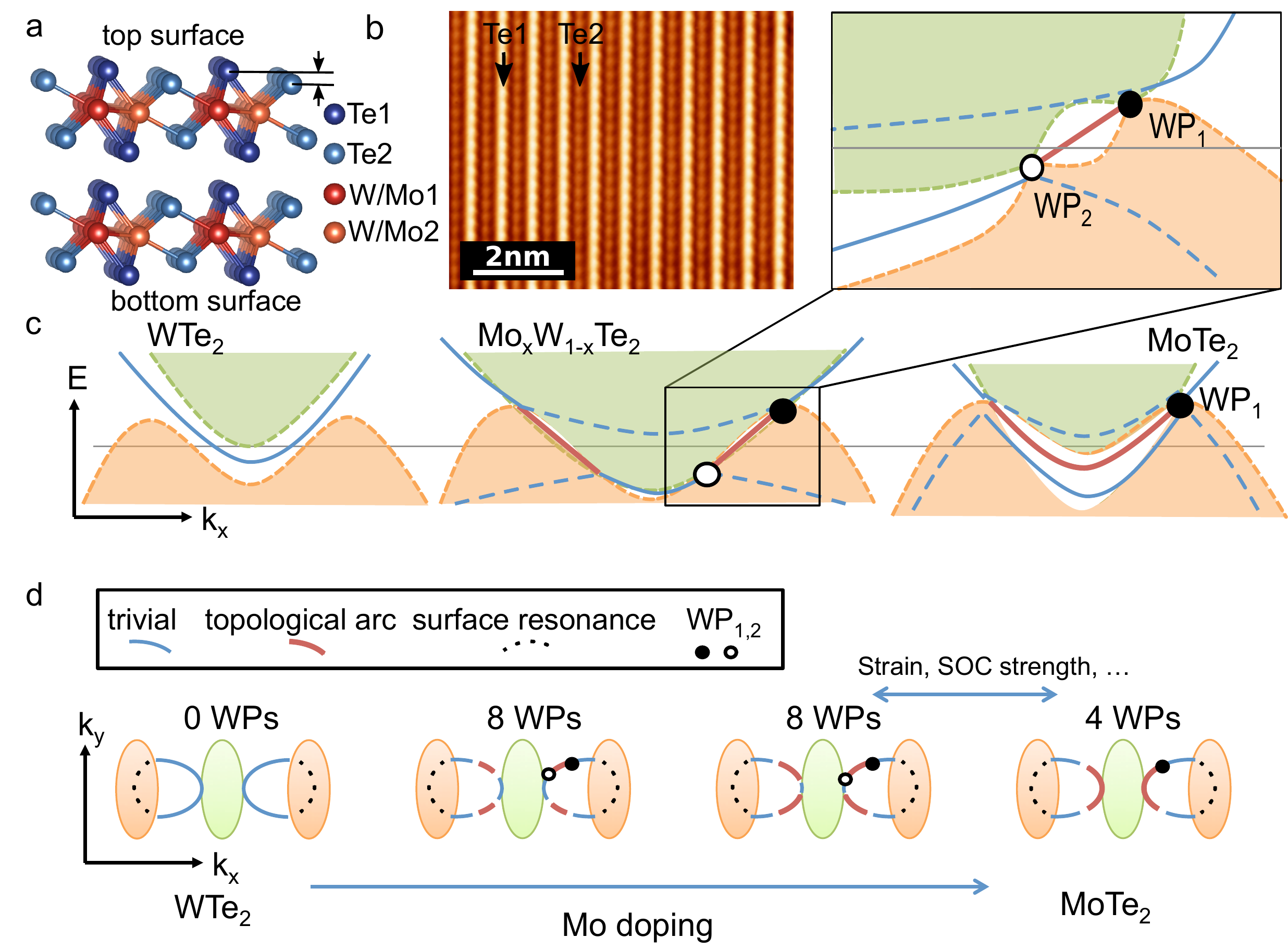}%
	\caption{{\bf a} Crystal structure of MoTe$_2$ and WTe$_2$;
	{\bf b} Atomically resolved STM topography of the WTe$_2$ surface. 
	The existence of two inequivalent Te sites labelled Te1 and Te2 with Te1 
	being slightly higher than Te2 gives rise to the column-like character visible in the STM image. 
	{\bf c} Schematic representation of the type-II Weyl phase diagram. 
	Although being topologically trivial, WTe$_2$ is found to lie very close to the topological phase transition. 
	Consequently, small lattice distortions can easily drive WTe$_2$ into a Weyl phase. 
	By starting from WTe$_2$ and substituting W with Mo, the system enters into a progressively more stable Weyl phase.
	{\bf d} Evolution of the topological Fermi arcs as a function of the Mo concentration. 
	By increasing the Mo concentration, topological Fermi arcs (red lines) become progressively larger. }	 
\label{Figure1}
\end{figure*}   

The Mo$_x$W$_{1-x}$Te$_2$ family is characterized by a layered structure 
which crystallizes in a $T_d$ orthorhombic cell lacking inversion symmetry 
at low temperatures (space group $Pmn$2$_1$) \cite {Clarke1978}. 
The transition metal (W and Mo) planes are separated by Te bilayers as shown in Fig.~1(a). 
Adjacent Te--(W/Mo)--Te trilayers  are weakly bound by van der Waals forces thus offering a natural cleaving plane.  
As a result, the surface exposed after cleaving is always Te-terminated. 
As illustrated in Fig.~1(a), the Te atoms occupy two inequivalent sites, 
labelled Te1 and Te2, with one slightly protruding over the other. 
This is reflected in the column-like character visible in the atomically resolved STM image reported in Fig.~1(b).

Density functional theory calculations indicated how, by starting from WTe$_2$---which lies close 
to a topological phase transition---and substituting W with Mo, the system enters 
into a progressively more stable Weyl phase as schematically illustrated in Fig.~1(c) \cite{CXC2016}. 
In particular, by increasing the Mo concentration, the Weyl points become well-separated in reciprocal space 
and thus cannot easily be annihilated by small perturbations \cite{CXC2016}. 
The larger Weyl point separation has direct consequences for the surface, 
with topological Fermi arcs getting progressively larger [see red lines in Fig.~1(d)]. 
In this respect, it is worth noticing that the surface electronic structure of these compounds 
is substantially complicated by the concomitant existence of trivial surface states. 
Although they do not form open arcs, trivial states partially overlap 
with the projected bulk electronic structure, giving rise to surface resonances [dashed lines in Fig.~1(d)] 
which are characterized by a reduced surface spectral weight. 
As a result, their pure surface state part (blue line) can effectively mimic an open arc-like contour.  

\begin{figure*}[t]   
	\includegraphics[width=.85\textwidth]{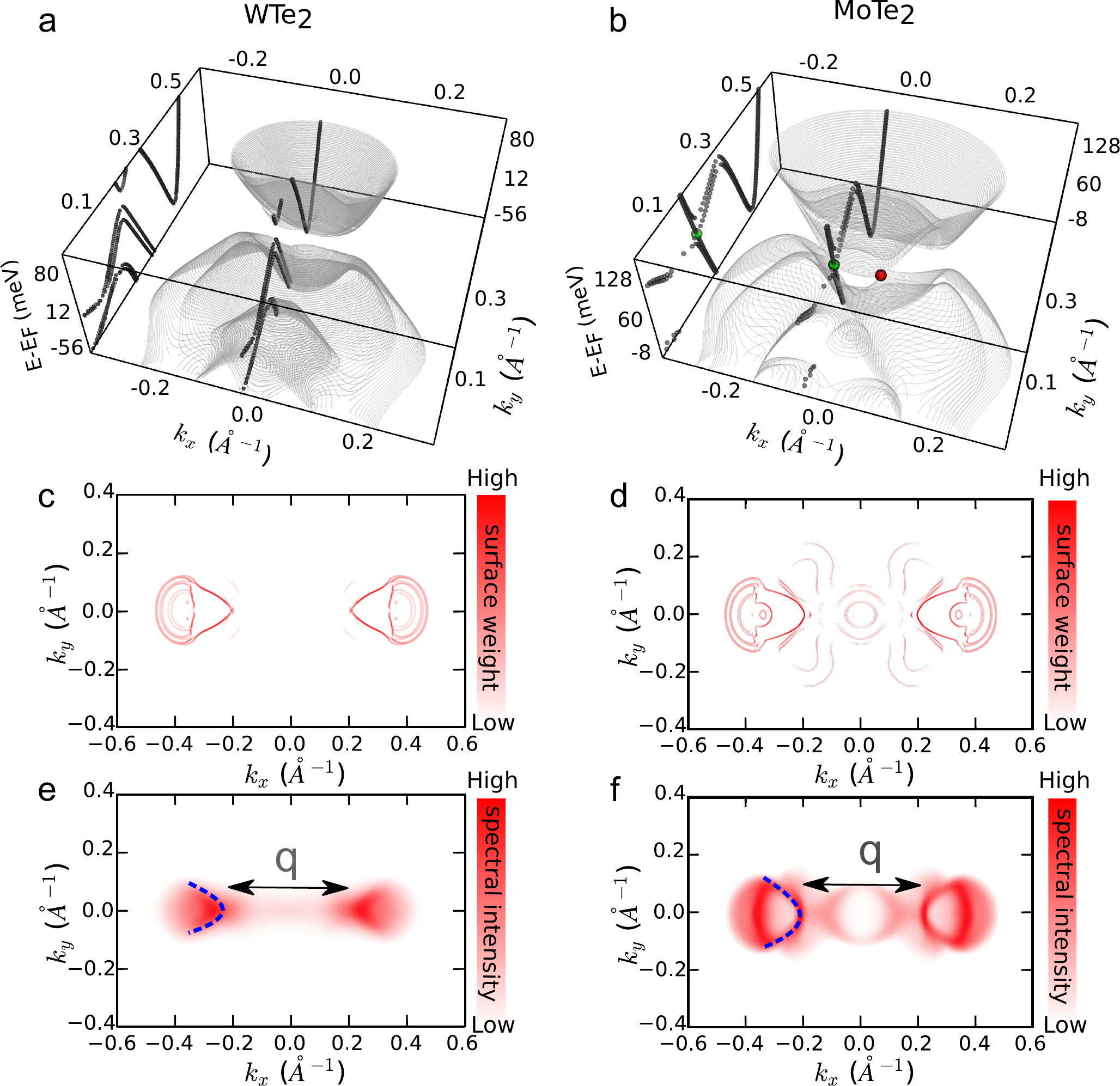}%
	\caption{{\bf a,b} Calculated bulk band structure at $k_z=0$, 
	{\bf c,d} theoretical and {\bf e,f} experimental Fermi surface for WTe$_2$ and MoTe$_2$, respectively. 
	In (a), the dotted lines show the band structure along $k_y$ at the particular $k_x$ 
	where the electron and hole pockets approach each other closest.
	In (b), the dotted lines show the band structure along $k_y$ at the specific $k_x$ 
	where Weyl points appear (green and red circles, reflection-symmetric around $k_x=0$). 
	The experimental Fermi surfaces have been obtained by angle-resolved photoemission spectroscopy 
	at a temperature $T < $ 60 K using photon energies of 20 and 24 eV for MoTe$_2$ and WTe$_2$, respectively. }	 
\label{Figure2}
\end{figure*}   

Experimentally investigating this phase diagram proved problematic. 
This is mainly due to two reasons: 
(i) Contrary to type-I Weyl materials, in type-II, the projection of the Weyl points onto the surface 
is overlapping with several bulk-derived trivial states, thereby complicating the discrimination 
of topological Fermi arcs from other states of trivial origin \cite{HMO2016,PhysRevX.6.031021,DWD2016,%
PhysRevB.94.121112,PhysRevB.94.121113,PhysRevB.94.161401,PhysRevB.94.241119,BSI2016}; 
(ii)  The Weyl points are expected to appear at energies above the Fermi level 
\cite{SGW2015,PhysRevB.92.161107,PhysRevLett.117.056805,CXC2016}, 
where they are inaccessible to conventional photoemission techniques \cite{HMO2016,PhysRevX.6.031021,DWD2016,%
PhysRevB.94.121112,PhysRevB.94.121113,PhysRevB.94.161401,PhysRevB.94.241119,BSI2016}. 
In this context, while a general consensus exists over the topological nature 
of the MoTe$_2$ arcs \cite{HMO2016,PhysRevX.6.031021,DWD2016}, 
the situation appears highly controversial for WTe$_2$. 
Although surface arcs have clearly been observed in several photoemission studies, 
their topological or trivial nature is highly debated with different studies reaching conflicting conclusions 
\cite{PhysRevB.94.121112, PhysRevB.94.121113,PhysRevB.94.161401,PhysRevB.94.241119}.  
$Ab$ $initio$ calculations also show that---while MoTe$_2$ is well inside the Weyl regime \cite{FN1}---%
WTe$_2$ is in close proximity to a Weyl phase transition \cite{PhysRevB.94.121112}. 
This can give rise to Weyl points of opposite chirality which are very close in reciprocal space and thus can easily be annihilated 
by very small lattice distortions induced by strain or temperature as discussed in Ref.~\onlinecite{PhysRevB.94.121112}. 
Therefore, it is particularly important to investigate the existence of universal signatures 
spanning the entire Weyl phase diagram, e.g.\ by comparing the two extreme cases, MoTe$_2$ and WTe$_2$.   

Here, we visualize the response of MoTe$_2$ and WTe$_2$ and discuss the results in terms of Weyl nodes and Fermi arcs. 
We identify the existence of a universal response of these systems to perturbations, which is found to be composition-independent. 
In particular, we report the emergence of well-defined quasiparticle interference patterns originating from surface arcs. 
Contrary to earlier studies \cite{DWD2016}: (i) we can clearly resolve their open contour 
and (ii) demonstrate that topological Fermi arc contributions are strongly suppressed 
because of their spin texture which protects them from back-scattering \cite{DWD2016,PhysRevLett.117.266804}. 
Furthermore, in line with the theoretical predictions of the response of Weyl semimetals, 
we reveal the emergence of new quasiparticles arising at the Weyl point energy, 
which lift the density of states minimum associated to the Weyl node \cite{PhysRevB.87.155123,doi:10.1080/00018732.2014.927109}. 

\begin{figure*}[t]   
	\includegraphics[width=.99\textwidth]{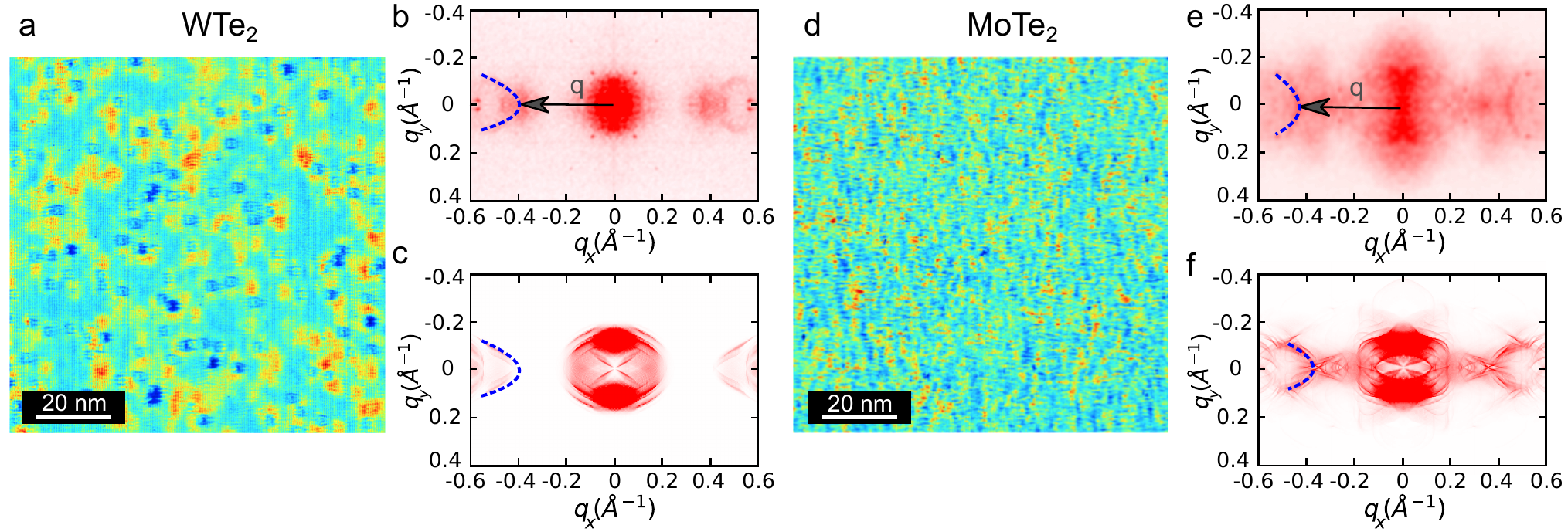}%
	\caption{{\bf a} Differential conductance d$I$/d$U$ map close to the Fermi ($E - E_{\rm F} = 10$\,meV); 
	{\bf b} Fourier-transformed d$I$/d$U$ map and 
	{\bf c} theoretically calculated quasi-particle interference pattern obtained on WTe$_2$. 
	{\bf d} Differential conductance d$I$/d$U$ map close to the Fermi ($E - E_{\rm F} = 10$\,meV); 
	{\bf e} Fourier-transformed d$I$/d$U$ map and 
	{\bf f} theoretically calculated quasi-particle interference pattern obtained on MoTe$_2$. 
	The quasi-particle interference pattern is dominated by scattering events between opposite Fermi arcs.  }	 
\label{Figure3}
\end{figure*}   

Fig.~2(a,b) report the electronic band structure of bulk WTe$_2$ and MoTe$_2$ at $k_z=0$, respectively, 
where due to the crystal symmetry Weyl points are possible \cite{SGW2015},
calculated with the density-functional-based full-potential relativistic Korringa-Kohn-Rostoker Green-function method 
\cite{Heers2011,Zimmermann2016,Long2014,Zeller2004}, respectively. 
Computational details can be found in supplementary material.
The combined effects of the inversion asymmetry of the crystal structure and the strong spin-orbit coupling 
characterizing these materials give rise to a spin-polarized band structure \cite{PhysRevB.94.195134}. 
However, whereas Weyl points are emerging in MoTe$_2$ 
[see green and red dots in Fig.~2(b) which identify one pair of Weyl points of opposite chirality], 
a gap between electron and hole pockets is clearly visible for WTe$_2$ [Fig.~2(a)], 
indicating the trivial character of this compound. 
In this context, it is worth noticing that the electronic structure of WTe$_2$ is very delicate. 
As discussed in Ref.\,\onlinecite{CXC2016}, small changes of the lattice constant can drive the system 
into a non-trivial state hosting 8 Weyl points, proving the close vicinity of a topological phase transition. 
Despite these differences highlighted by band structure calculations, 
our angle-resolved photoemission data reported in Fig.~2(e,f) 
reveal the presence of arc-like electronic structures in both compounds (see dashed lines). 
An overall agreement between theory and experiments is found. 
Although it is tempting to assign the arc-like features in Fig.~2(e,f) to topological Fermi arc states, 
comparison with calculated constant-energy contours reported in Fig.~2(c,d) reveal a more complicated scenario.  
Indeed, $ab$ $initio$ calculations reveal that several electronic features coexist 
within a very small part of the first Brillouin zone which cannot fully be resolved 
by photoemission experiments \cite{FN2}.    
It is worth stressing that, as discussed above, an analysis of our calculations reveals that most of the arcs 
visible in Fig.~2(c,d) are of trivial origin and that only for MoTe$_2$ also topological Fermi arcs are present.  
This proves that the observation of surface arcs is a necessary, 
but not a sufficient condition to unequivocally imply the existence of Weyl points.

\begin{figure*}[t]   
	\includegraphics[width=.99\textwidth]{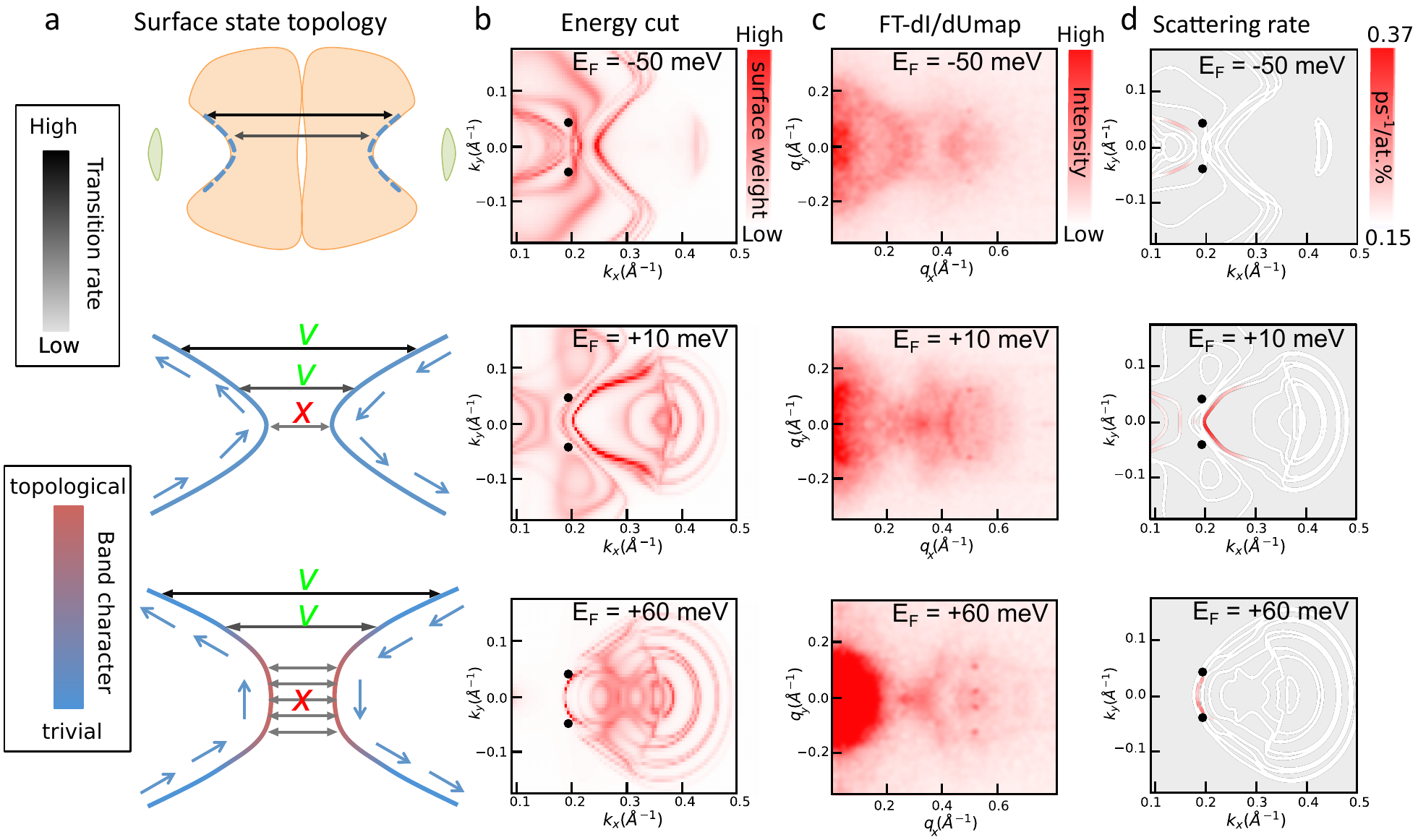}%
	\caption{{\bf a} Schematic representation of the scattering events among Fermi arcs in MoTe$_2$, 
	{\bf b} theoretically calculated surface-projected constant energy cuts, 
	{\bf c} experimentally obtained Fourier transformed quasi particle interference patterns and 
	{\bf d} theoretically calculated scattering rate. In (b-d) the black dots identify the Weyl points. 
	All results are given for three representative energies:  below the Fermi (top panels), 
	close to the Fermi (middle panels), close to the position of the Weyl points, 
	where the extension of the topologically non-trivial arcs is maximized (bottom panels). }	 
\label{Figure4}
\end{figure*}   

\begin{figure*}[t]   
	\includegraphics[width=.85\textwidth]{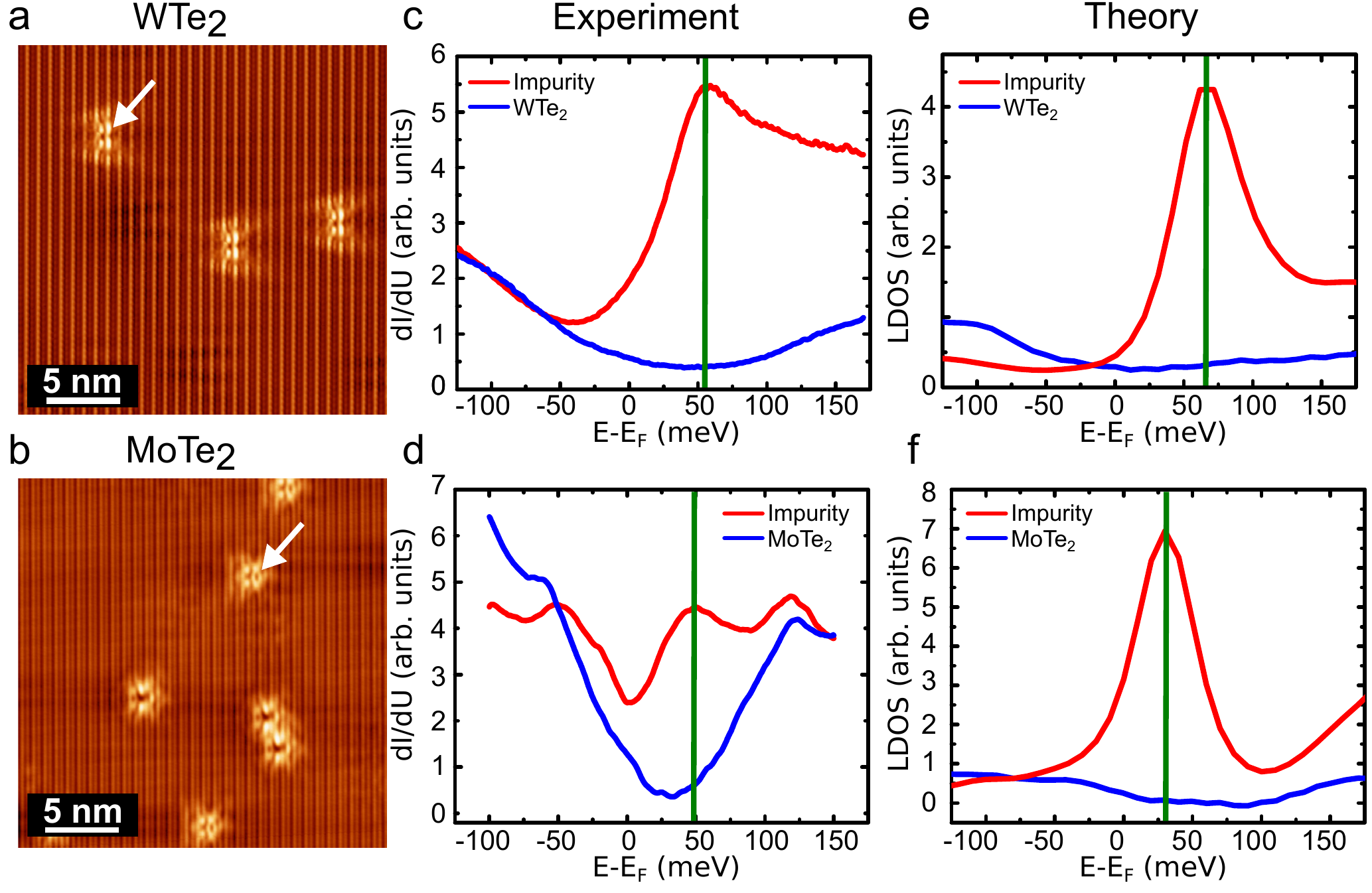}%
	\caption{{\bf a,b} Topographic images acquired on WTe$_2$ and MoTe$_2$. 
	In both cases, intrinsic defects are present on the surface. 
	{\bf c,d} Comparison of scanning tunneling spectroscopy data taken on a defect-free area (blue line) 
	and by positioning the tip on top of antisites defects (W and Mo substituting Te) 
	revealing the emergence of quasiparticle resonances close the Weyl point energy (see discussion in the text). 
	{\bf e,f}  {\em Ab-initio} calculated local density of states on the pristine surface (blue line) 
	and on top of an antisite defect (red line) confirm the experimental findings. 
	Green vertical lines are used as marker to identify the position of the peak maximum.  }	 
\label{Figure5}
\end{figure*}   
 
To disentangle trivial from topological Fermi arcs contributions, we performed quasi-particle interference experiments. 
This technique makes use of the standing-wave pattern generated by elastic scattering of electronic states 
at surface defects and has been proven to be a powerful method to test the properties 
of topological materials \cite{RSP2009,PhysRevLett.103.266803,SRB2014,ZOH2014,JZG2014,PhysRevB.95.035114}. 
Contrary to conventional photoemission spectroscopy, it gives access to both occupied and unoccupied electronic states 
thereby providing a complete spectroscopic characterization of all relevant electronic features around the Fermi energy. 
This is particularly important for type-II Weyl materials, where Weyl points are theoretically expected to emerge 
above the Fermi level \cite{SGW2015,PhysRevB.92.161107,PhysRevLett.117.056805,CXC2016}. 
Fig.~3(a) and (d) report differential conductance d$I$/d$U$ maps 
acquired in close proximity of the Fermi level ($E - E_{\rm F} = 10$\,meV) on WTe$_2$ and MoTe$_2$, respectively. 
Their Fourier transformations---reported in panels (b) and (e)---allow to conveniently analyze scattering channels in reciprocal space. 
Our results reveal the emergence of clear arc-like interference patterns 
as indicated by the black arrows in both compounds which develop 
along the $q_x$ direction at approximately $q_x = 0.4$ \AA~(see dashed blue lines). 
To shed light on the origin of these vectors, experimental data have been compared with calculated interference patterns.  
Our theoretical approach employs the extended-JDOS model \cite{PhysRevB.94.075137}, 
with the impurity's electronic structure and scattering amplitude calculated by means KKR method. 
Results for WTe$_2$ and MoTe$_2$ are reported in Fig.~3(c) and (f). 
As for the experimental results, an arc-like feature is clearly visible in both cases along the $q_x$ direction (see dashed lines), 
which originates from intra-arc scattering among opposite Fermi arcs. 
This assignment is further supported by direct comparison with photoemission data [cf.\ Fig. 2(e,f), 
where the scattering vector $q$ corresponds to the distance connecting opposite arcs]. 
Indeed, momentum conservation requires $k_f = k_i + q$, where $k_i$ and $k_f$ are the wave vectors 
of initial and final states and $q$ is the scattering vector connecting them. 
A quantitative comparison with the experimental constant energy contours reported in Fig.~2(e,f) 
allows to directly link these interference phenomena to intra-arc scattering among opposite Fermi arcs. 
Furthermore, contrary to photoemission data, the higher surface sensitivity of STM 
unequivocally proves the open contour character of the arcs.  

In this context, it is worth noticing that the close proximity of WTe$_2$ to a Weyl transition 
allows one to safely exclude any significant contribution of topological Fermi arcs 
to the observed interference patters for this compound. 
Even when considering a slightly distorted structure hosting Weyl points, 
they would be so close in reciprocal space that the extension of topological Fermi arcs connecting them would be negligible. 
This is not the case in MoTe$_2$ where---by progressively moving towards the energy position 
of the Weyl points---topologically non-trivial Fermi arcs span a much larger fraction of the Brillouin zone.  
This is illustrated by the schematic representation reported in Fig.~4(a) 
and the theoretically calculated constant energy cuts displayed in Fig.~4(b). 
As shown in panel (a), by progressively increasing the energy (from top to bottom panel) 
the trivial part (blue line) of the arc is absorbed into the bulk electronic band structure
whereas the topologically non-trivial arc (red line) dominates the scene. 
This transition from trivial to topology-dominated Fermi arcs is experimentally investigated in Fig.~4(c).
At energies well below the Weyl points (upper panel)  only trivial arcs exist.  
As a consequence, the experimental data show only weak interference patterns. 
This can be traced back to the combined effect of (i) the V-shape visible in the calculated constant energy contour 
which results in poor nesting conditions and (ii) the overlap with projected bulk pockets 
making these states surface resonances that are much less localized at the surface 
than ``real'' surface states [cf.\ top panel of Fig.~4(a)]. 
By going up in energy, the surface states become well separated from bulk states [cf.\ middle panel of Fig.~4(a)] 
while simultaneously getting progressively larger, and thus occupying a larger fraction of the Brillouin zone. 
As a result, a well-defined arc-like interference pattern appears (see middle panel) 
which---according to calculations---is dominated by trivial states. 
By moving to even higher energies we move closer to the Weyl points 
and the topological Fermi arcs prevail [cf.\ bottom panel of Fig.~4(a)].  
This process is associated to a significant flattening of the arc, where several equivalent vectors can connect 
opposite parallel segments of the Fermi arcs [see grey arrows in Fig.~4(a), lower panel]. 
Despite this scenario supporting optimal nesting conditions, the scattering intensity drops (see bottom panel). 
Comparison with $ab$ $initio$ calculations reveals that this effect 
is directly linked to the spin texture of the topological Fermi arcs. 
Indeed, as schematically illustrated by the blue arrows in Fig.~4 (a), at $E - E_{\rm F} = +60$~meV
the spin polarization is basically pointing in opposite directions for opposite topological Fermi arcs segments 
(spin-resolved constant-energy contours are reported in the supplementary material). 
This spin texture results in an effective protection against scattering. 
Such a behavior is reminiscent of the forbidden backscattering 
originally discussed in Rashba systems \cite{PhysRevLett.93.196802} and, 
more recently, in topological insulators \cite{PhysRevLett.103.266803}. 
In the present case, however, the presence of large parallel segments with opposite spin polarization 
significantly extend the protection  well beyond time-reversal symmetry partner states.
   
These conclusions, which are based on a combined analysis of the spin-resolved band structure 
and quasi-particle interference mapping, are further quantitatively supported 
by the calculated scattering rates reported in Fig.~4(d). 
By progressively increasing the energy, the surface arcs occupy a larger fraction of the Brillouin zone. 
Consequently, more scattering vectors connecting opposite arcs become possible 
and the scattering intensity rises (see red line in the middle panel). 
However, once the topological Fermi arcs set in (lower panel) the scattering intensity drops 
because of the discussed spin-texture protection mechanism.

Finally, we discuss the effect impurities have on Weyl nodes. 
Recent theoretical predictions showed that a common characteristic of Dirac-like materials 
is the emergence of impurity-induced quasiparticles which lift the Dirac node \cite{doi:10.1080/00018732.2014.927109}. 
This behavior has been recently confirmed in topological insulators, 
where impurity resonances induced by magnetic dopants have been shown to effectively fill the gap 
which is expected to open at the Dirac point in magnetically ordered samples \cite{SBB2016}.  
A similar behavior has been proposed to arise in Weyl semimetals.  
In particular, scattering at localized impurities is expected to lift the Weyl node by inducing new quasiparticle resonances 
close to the Weyl point energy \cite{PhysRevB.87.155123,doi:10.1080/00018732.2014.927109}. 
The emergence of these quasiparticles has been theoretically proposed as a signature of a Weyl phase. 
This has been experimentally investigated in Fig.~5, where panels (a) and (b) report topographic images 
acquired on the two different compounds, i.e., WTe$_2$ and MoTe$_2$. 
Intrinsic defects highlighted by arrows are visible in both cases. 
They have been identified as anti-sites (W or Mo atoms substitute Te in the topmost layer) 
and are commonly found in transition metal dichalcogenides \cite{584907111,HZP2015,doi:10.1021/acsnano.5b05854}. 
Fig.~5(c,d) reports STS data acquired on both materials by positioning the tip on top (red line) and far away from a defect (blue line). 
The minimum, visible in proximity to the Fermi level on defect-free areas, highlights the semimetallic character of these compounds.  
However, on top of defects a strong peak is visible in both materials which lifts the local density-of-states minimum. 

These defect-induced quasiparticle resonances appear very close to the energy 
where Weyl points are expected to emerge (see vertical green line which highlights the peak maximum). 
These findings are supported by the {\it ab initio}-calculated local-density-of-states at the impurity atom
reported in Fig.~5(e) and (f) for WTe$_2$ and MoTe$_2$, respectively. 
Both reveal the emergence of quasiparticle resonances located at energies which are in excellent agreement with the experiments. 
In particular, in MoTe$_2$ the experimentally observed impurity-induced quasiparticle resonance 
is positioned at the calculated Weyl point energy ($E - E_{\rm F}= +48$~meV).  
As discussed above, for WTe$_2$ our calculations predict a trivial material near to a topological phase transition into a Weyl phase. 
In this case, the impurity-induced resonances emerge close to the very narrow energy gap 
between electron and hole pockets, at an energy where Weyl points are expected to emerge 
according to Refs.~\onlinecite{SGW2015,PhysRevB.94.121112,PhysRevB.94.241119}. 
This observation provides strong evidence that topological Weyl transitions 
are continuous smooth transitions of the global bulk band structure. 
It follows that, although topological indexes change when driving the system through a topological quantum phase transition, 
the band structure-dependent experimental observables---such as the impurity-induced resonances 
reported here---can be continuously tuned, and are not characterized by any discontinuity, i.e., on/off behavior. 
In this sense, Weyl phase transitions appear to behave similarly to topological insulators phase transitions where, 
by approaching the critical point at which the bulk band structure becomes inverted, 
spin-polarized helical surface states progressively emerge within the bulk gap \cite{XNB2015}.

We would like to stress that the relevance of our observations
  goes well beyond topological band-structure aspects. It is
  long known that disorder, and especially resonant impurities, can significantly impact onto
  transport properties. Even recently, the presence of defects has been invoked to be at
  the origin of both positive as well as negative magnetoresistance
  effects in topological semimetals \cite{PhysRevLett.114.117201,Li2017}.  In this context, our
  observations contribute by providing a detailed microscopic picture
  of the resonant scattering off impurities in type-II Weyl
  semimetals. In particular, we demonstrate that intrinsic defects
  significantly alter the local density of states close to the Weyl
  points, ultimately changing the low-energy spectrum of Weyl
  semimetals. We conclude that the presence of defects cannot be
  overstressed and suggest that they play an important role in
  determining the fascinating transport properties of this class of
  materials  \cite{MGS2014,YEH2016}.

Overall, we reveal the existence of a universal response of the type-II Weyl semimetals phase diagram. 
We show that surface arcs dominate the interference pattern, 
with the topological Fermi arc contribution being strongly suppressed by its spin texture. 
In agreement with theoretical predictions, we also demonstrated 
that impurity-induced quasiparticle resonances emerge close to the energy where Weyl points are expected. 
Our observations highlight that the functional response of both surface and bulk states to perturbations 
in this class of materials does not depend on whether we have passed the Weyl phase transition or we are simply close to it. 
This allows to infer the existence of a stoichiometry-independent response to perturbations for type-II Weyl, 
providing a unifying picture of the type-II Weyl phase diagram.  

This work was supported by the Deutsche Forschungsgemeinschaft 
within SPP 1666 (Grant No. BO1468/21-1 and MA4637/3-1)
and through SFB 1170 ``ToCoTronics'' (project A02).  
P.R., P.M., and S.B. acknowledge financial support from the VITI project of the Helmholtz Association 
as well as computational support from the JARA-HPC Supercomputing Centre at RWTH Aachen University.

\bibliography{Weyl_bib}

\begin{thebibliography}{10}
\bibitem{FN1}
Note that the possible phase transition between 4 or 8 wps (as indicated in
  fig.1, bottom right) will not play a role here since still there will be a
  very large topological fermi arc.
\end{thebibliography}
\bibliographystyle{unsrt}

\end{document}